\documentclass[aps,prc,preprint,showpacs,floatfix]{revtex4-1}  
\usepackage{graphicx}
\usepackage{amsmath}
\usepackage{color}

\newcommand {\nc} {\newcommand}
\nc {\beq} {\begin{eqnarray}} \nc {\eol} {\nonumber \\} \nc {\eeq}
{\end{eqnarray}} \nc {\eeqn} [1] {\label{#1} \end{eqnarray}} \nc
{\eoln} [1] {\label{#1} \\} \nc {\ve} [1] {\mbox{\boldmath $#1$}}
\nc {\rref} [1] {(\ref{#1})} \nc {\Eq} [1] {Eq.~(\ref{#1})} \nc
{\re} [1] {Ref.~\cite{#1}} \nc {\dem} {\mbox{$\frac{1}{2}$}} \nc
{\arrow} [2] {\mbox{$\mathop{\rightarrow}\limits_{#1 \rightarrow
#2}$}}
\begin{document}
\title{Astrophysical $S$ factor and rate of $^{7}{\rm Be}(p, \gamma)^{8}{\rm B}$
direct capture reaction in a potential model}

\author {E. M. Tursunov}
\email{tursune@inp.uz} \affiliation {Institute of Nuclear Physics,
Academy of Sciences, 100214, Ulugbek, Tashkent, Uzbekistan}
\author {S. A. Turakulov}
\email{turakulov@inp.uz} \affiliation {Institute of Nuclear Physics,
Academy of Sciences, 100214, Ulugbek, Tashkent, Uzbekistan}
\author{A. S. Kadyrov}
\email{a.kadyrov@curtin.edu.au} \affiliation{Curtin Institute for
Computation and Department of Physics and Astronomy, Curtin
University, GPO Box U1987, Perth, WA 6845, Australia}
\author{L. D. Blokhintsev}
\email{blokh@srd.sinp.msu.ru} \affiliation{Skobeltsyn Institute of
Nuclear Physics, Lomonosov Moscow State University, Russia}

\begin{abstract}
The astrophysical $^7{\rm Be}(p, \gamma)^8{\rm B}$ direct capture
process is studied in the framework of a two-body single-channel
model with potentials of the Gaussian form.
A modified potential is
constructed to reproduce the new  experimental value of the $S$-wave
scattering length and the known astrophysical $S$ factor at the
Gamow energy, extracted from the solar neutrino flux. The resulting
potential is consistent with the theory developed by Baye [Phys. Rev. C {\bf
62} (2000) 065803] according to which the $S$-wave scattering length and the astrophysical
$S$ factor at zero energy divided by the square of ANC are related.
The obtained results for the astrophysical $S$ factor at
intermediate energies are in good agreement with the two data sets
of Hammache {\it et al.} [Phys. Rev. Lett. {\bf 86}, 3985 (2001);
{\it ibid.} {\bf 80}, 928 (1998)]. Linear extrapolation to zero
energy yields $ S_{17}(0) \approx (20.5 \pm 0.5) \, \rm eV \, b $,
 consistent with the Solar Fusion II estimate. The
calculated reaction rates are substantially lower than the results
of the NACRE II collaboration.
\end{abstract}


\maketitle

\section{Introduction}

The astrophysical capture process $^{7}{\rm Be}(p, \gamma)^{8} {\rm
B}$ is the most important nuclear reaction of the pp-chain in the
Solar Fusion Model and in stellar nucleosynthesis
\cite{rol88,ang99,adel11,fiel11}. A realistic estimate of the
reaction rate of this process is crucial for the solution of the
solar neutrino problem. The core temperature of the Sun can be
determined through the measurements of the $^8$B neutrino flux with
a precision  of about 9$\%$ \cite{sno04}. The rate of the $^{7}{\rm
Be}(p, \gamma)^{8}{\rm B}$ reaction is used for modeling this solar
neutrino flux.
 \par Many original research papers have
been published in addition to reviews \cite{ang99,adel11}. Direct
measurements face difficulties due to large Coulomb forces at low
energies \cite{fil83,moto94,baby03,jun03,ham98,ham01,jun10}. Coulomb
dissociation of $^8$B in the field of a heavy target have been
experimentally studied in Refs.
\cite{schum03,baur86,iwa99,dav01,schum03,schum06}. However, none of these
experimental studies could reach the energies below the solar Gamow
window at 0.019 MeV. As a result, the extrapolated value $S_{17}(0)$
of the astrophysical $S$ factor in the "Solar Fusion II" (SF II)
workshop
 \begin{equation}
  S_{17}(0)=(20.8\pm0.7_{expt} \pm 1.4_{theor}) \, \rm {eV \, b}
  \label{eq01}
  \end{equation}
 has a large uncertainty \cite{adel11}.
\par From the theory point of view, potential models
\cite{rob73,typ97,dav03,dub19}, R-matrix parametrization
\cite{bar95}, microscopic models \cite{desc94,csy95,desc04},
three-body model \cite{gri98}, {\it ab-initio} calculations
\cite{nav06,nav11}, Skyrme-Hartree-Fock theory \cite{chan03} and
halo effective-field theory \cite{halo15} have been developed.
Results of most theoretical studies for $S_{17}(0)$ belong to the
aforementioned uncertainty range of the SF II estimate. In Ref.
\cite{azha99} the reaction
$^{10}$B($^7$Be,$^8$B)$^9$Be
was used for extracting the asymptotic normalization
coefficient (ANC) $C$ for the the virtual transition $p+^7${Be}
$ \to \, ^8${B}. Similarly, in Ref. \cite{olim16} the ANC was
extracted from the data of the $^7\rm Be(d,n)^8\rm B$ transfer
reaction. In Ref. \cite{tra03} the ANC was derived from the
experimental cross section of the $^{13}\rm {C}(^7\rm Li,^8\rm
Li)^{12}\rm C$ charge-conjugate reaction. As first established in
Ref. \cite{mukh94}, the astrophysical $S$ factor at low energies is
mainly determined by the ANC. The idea is widely used for
estimating the astrophysical $S$ factor of capture reactions
\cite{azha99,mukh16,olim16}.

Besides the value of the astrophysical $S$ factor at zero energy,
the most important property is the energy dependence of the
astrophysical $S$ factor at low energies below the Gamow window. In
Ref. \cite{baye00a} the coefficients of the Taylor expansion in
terms of energy around zero energy have been found for a given
potential. Due to the fact that the largest contribution to the
astrophysical $S$ factor of the process at low energies comes from
the initial $S$-wave $p+^7\rm Be$ scattering state, in
Ref. \cite{baye00b} the dependence of $S(0)/C^2$ and $S'(0)/S(0)$ on
the $S$-wave $p+^7\rm Be$ scattering length have been studied in
details and important formulas have been derived.

On the other hand, recently in Ref. \cite{pan19} the most precise
experimental values $a_{01}=17.34^{+1.11}_{-1.33}$ fm and
$a_{02}=-3.18^{+0.55}_{-0.50}$ fm for the $s$-wave scattering
lengths have been obtained in the spin=1 and spin=2 channels,
respectively. Additionally, a new data for the astrophysical $S$
factor at the Gamow energy has been extracted from the solar
neutrino flux \cite{tak18} to be
\begin{equation}
 S_{17}(19^{+6}_{-5} \,\, {\rm keV})=(19.0\pm 1.8) \, \rm {eV \,b}.
 \label{eq3}
\end{equation}
The aim of the present study is to estimate $S_{17}$ and
corresponding reaction rates in the potential model which reproduces
new values of the $S$-wave scattering length and of $S_{17}$ at the
Gamow energy. This work is based on a single-channel potential model
\cite{dub19}. First we examine and optimize the $S$-wave
potential parameters by fitting to the new value of $a_{01}$, then we fit
the bound $^3P_2$ state potential parameters based on the new values of $S_{17}$ at the Gamow energy found in Ref.
\cite{tak18}, as described above. Then consistency of the resulting potential with the theory
of Ref. \cite{baye00b} is examined.

In Section 2 the theoretical model is described. Section 3 contains
the numerical results. Conclusions are drawn in the last section.

\section{Theoretical model}
\subsection{Wave functions}

 In the single-channel potential model \cite{tur15,tur18,tur21}, the initial and final state wave functions are defined as
\begin{eqnarray}
\Psi_{lS}^{J}=\frac{u_E^{(lSJ)}(r)}{r}\left\{Y_{l}(\hat{r})\otimes\chi_{S}(\xi)
\right\}_{J M}
\end{eqnarray}
and
\begin{eqnarray}
\Psi_{l_f S'}^{J_f}
=\frac{u^{(l_fS'J_f)}(r)}{r}\left\{Y_{l_f}(\hat{r})\otimes\chi_{S'}(\xi)
\right\}_{J_f M_f},
\end{eqnarray}
respectively. The initial $p -^7${\rm Be} scattering states in the
$^3S_1$, $^3P_0$, $^3P_1$, $^3P_2$, $^3D_1$, $^3D_2$, $^3D_3$,
$^3F_3$  partial waves are described by the radial wave functions
which are solutions of the two-body Schr\"{o}dinger equation
\begin{align}
\left[-\frac{\hbar^2}{2\mu}\left(\frac{d^2}{dr^2}-\frac{l(l+1)}{r^2}\right)+V^{
lSJ}(r)\right] u_E^{(lSJ)}(r)=E u_E^{(lSJ)}(r),
\end{align}
where $\mu$ is the reduced mass of $p$ and $^7\rm Be(3/2-)$,
${1}/{\mu}={1}/{m_1}+{1}/{m_2}$ , and $V^{lSJ}(r)$ is a two-body
potential in the partial wave with the orbital angular momentum $l$, spin
$S$ and total angular momentum $J$. The wave function $u^{(l_fS'J_f)}(r)$ of
the final $^3P_2$ ground state is calculated as a solution of the
bound-state Schr\"{o}dinger equation. The Schr\"{o}dinger equation
is solved using the Numerov algorithm. The cross section, the
astrophysical $S$ factor and the reaction rates are estimated using
the accurate wave functions of the initial and final states.
The initial scattering wave function is found subject to the
asymptotic condition
\begin{align}
 u_E^{(lSJ)}(r) \arrow{r}{\infty}
\cos\delta_{lSJ}(E) F_l (\eta,kr) + \sin\delta_{lSJ}(E)
G_l(\eta,kr), \label{eq220} \end{align}
where $k$ is the wave number of the
relative motion, $\eta$ is the Zommerfeld parameter, $F_l$ and $G_l$
are regular and irregular Coulomb functions, respectively, and
$\delta_{lSJ}(E)$ is the phase shift in the $(l,S,J)$th partial
wave.

The $p -^7${\rm Be} two-body potential has the Gaussian form
\cite{dub19}:
 \beq
 V^{lSJ}(r)=V_0 \exp(-\alpha_0 r^2)+V_c(r),
\label{pot}
 \eeq
where the Coulomb part is taken in a point-like potential form
\cite{dub19}.
\subsection{Cross sections of the radiative-capture process}
The cross sections for radiative-capture process can be expressed as
\cite{ang99, dub19}
\begin{eqnarray}
\sigma(E)=\sum_{J_f \lambda \Omega}\sigma_{J_f \lambda}(\Omega),
\end{eqnarray}
where $\Omega=$ E  or M (electric or magnetic transition), $\lambda$
is a multiplicity of the transition, $J_f$ is the total angular
momentum of the final state. For a particular final state with total angular
momentum $J_f$ and multiplicity $\lambda$ we have \cite{ang99}
\begin{align}
 \sigma_{J_f \lambda}(\Omega) =& \sum_{J}\frac{(2J_f+1)} {\left
[S_1 \right]\left[S_2\right]} \frac{32 \pi^2 (\lambda+1)}{\hbar
\lambda \left( \left[ \lambda
\right]!! \right)^2} k_{\gamma}^{2 \lambda+1} C^2(S) \nonumber \\
&\times \sum_{l S}
 \frac{1}{ k_i^2 v_i}\mid
 \langle \Psi_{l_f S'}^{J_f}
\|M_\lambda^\Omega\| \Psi_{l S}^{J} \rangle \mid^2,
\end{align}
where $l$ and $l_{f}$ are the orbital momenta of the initial and final
states, respectively; $k_i$ and $v_i$ are the wave number and
speed of the $p - ^7${\rm Be} relative motion in the entrance
channel, respectively; $S_1$ and $S_2$ are spins of the clusters $p$
and $^7${\rm Be}, $k_{\gamma}=E_\gamma / \hbar c$ is the wave number
of the photon corresponding to energy $E_\gamma=E_{\rm th}+E$, where
$E_{\rm th}$ is the threshold energy for the breakup reaction
$\gamma+^8{\rm B} \to ^7{\rm Be}+p$. Constant $C^2(S)$ is a
spectroscopic factor \cite{ang99}. Within the potential approach
where the bound and scattering properties (energies, phase shifts
and scattering length) are reproduced, a value of the spectroscopic
factor must be taken equal to 1 \cite{mukh16}. We also use
short-hand notations $[S]=2S+1$ and $[\lambda]!!=(2\lambda+1)!!$.

The reduced matrix elements are evaluated between the initial
$\Psi_{l S}^{J}$ and final $\Psi_{l_f S'}^{J_f}$ state wave
functions. The electric transition operator in the long-wavelength
approximation reads as
\begin{eqnarray}
M_{\lambda \mu}^{\rm E}=e \sum_{j=1}^{A}
Z_j{r'_j}^{\lambda}Y_{\lambda \mu}(\hat{r'}_j),
\end{eqnarray}
where $\vec {r'}_{j}= \vec{r}_j-\vec{R}_{cm}$ is the position
of the $j$th particle in the center of mass system. Its reduced
matrix elements can be evaluated as \cite{ang99}
\begin{eqnarray}
\langle \Psi_{l_f S'}^{J_f}\|M_\lambda^{\rm E}\| \Psi_{l S}^{J} \rangle
 &=& e\left[Z_1 \left( \frac{A_2}{A} \right)^{\lambda}+Z_2
\left(\frac{-A_1}{A} \right)^{\lambda} \right]   \delta_{S S'}  \\
\nonumber && \times (-1)^{J+l+S}\left(\frac{[\lambda][l][J]}{4
\pi}\right)^{1/2} C^{l_f 0}_{\lambda 0 l 0} \left\{
\begin{array}{ccc}
J & l & S \\
l_{f} & J_{f} & \lambda
\end{array} \right\} \int^{\infty}_{0} u_{E}^{(lSJ)}(r)r^{\lambda} u^{(l_fSJ_f)} (r) dr,
\end{eqnarray}
where $A_1$, $A_2$  are the mass numbers of the clusters in the entrance
channel, $A=A_1+A_2$.
The magnetic transition operator reads as \cite{ang99}
\begin{eqnarray}
M_{1 \mu}^{\rm M}&=& \sqrt{\frac {3}{4 \pi}} \sum_{j=1}^{A}
\left[\mu_N
\frac{Z_j}{A_j}\hat{l}_{j \mu} + 2 \mu_j \hat{S}_{j \mu} \right] \\
\nonumber & = & \sqrt{\frac {3}{4 \pi}}\left[\mu_N \left( \frac{A_2
Z_1}{A A_1} + \frac{A_1 Z_2}{A A_2} \right) \hat{l}_{r
\mu}+2(\mu_1\hat{S}_{1\mu}+\mu_2\hat{S}_{2\mu})\right],
\end{eqnarray}
where $\mu_N$ is the nuclear magneton, $\mu_j$ is the magnetic
moment and $\hat{l}_{j \mu}$ ($\mu=-1,0,+1$) is the projection of
the orbital angular momentum of $j$th particle. The projection of
the orbital angular momentum of the relative motion is denoted as
$\hat{l}_{r \mu}$.
  The magnetic M1 transition
operator consists of the orbital and spin parts:
\begin{eqnarray}
M_{1 \mu}^{\rm M}=\sqrt{\frac{3}{4\pi}} \left[
M_1(\emph{l})+M_1(\emph{S})\right].
\end{eqnarray}
The orbital part of the reduced matrix elements of the magnetic M1
transition operator reads as
\begin{eqnarray}
\langle \Psi_{l_f S'}^{J_f}\|M_1(\emph{l})\| \Psi_{l S}^{J} \rangle
& = & \mu_N \left( \frac{A_2 Z_1}{A A_1} + \frac{A_1
 Z_2}{A A_2} \right) \sqrt{l(l+1)[J][l]} \\
 \nonumber && \times (-1)^{\kappa_1}
\left\{
\begin{array}{ccc}
l & S & J_f \\
J & 1 & l
\end{array}
\right\} \delta_{l l_f} \delta_{S S'} I_{if},
\end{eqnarray}
where the exponential part of the phase factor $\kappa_1=S+1+J+l$.
 The spin part of the magnetic M1 transition
operator for the first particle (proton)
\begin{eqnarray}
\langle \Psi_{l_f S'}^{J_f}\|M_1^{\rm M}(S_1)\| \Psi_{l S}^{J} \rangle & =
& 2 \mu_p(-1)^{\kappa_2} \sqrt{S_{1}(S_{1}+1)[S_1][S][S'][J]} \\
\nonumber && \times \left\{
\begin{array}{ccc}
S_1 & S_2 & S \\
S' & 1 & S_1
\end{array}
\right\} \left\{
\begin{array}{ccc}
S & l & J \\
J_f & 1 & S'
\end{array}
\right\} \delta_{l l_f} I_{if},
\end{eqnarray}
with the exponential part of the phase factor
$\kappa_2=S_1+S_2+2S+l+J_f$. In the above formula and everywhere we
set $S_1=S_p$=1/2, $S_2=S(^7$Be)=3/2 and $S'=S=$1 due to the use of
the single-channel approximation. The spin part of the reduced
matrix elements of the M1 transition operator for the second
particle ($^7$Be) reads as
\begin{eqnarray}
\langle \Psi_{l_f S'}^{J_f}\|M_1^{\rm M}(S_2)\| \Psi_{l S}^{J} \rangle & =
& 2 \mu_{^7 Be}(-1)^{\kappa_3} \sqrt{S_{2}(S_{2}+1)[S_2][S][S'][J]} \\
\nonumber & & \times \left\{
\begin{array}{ccc}
S_2 & S_1 & S \\
S' & 1 & S_2
\end{array}
\right\} \left\{
\begin{array}{ccc}
S & l & J \\
J_f & 1 & S'
\end{array}
\right\} \delta_{l l_f} I_{if},
\end{eqnarray}
where $\kappa_3=S_1+S_2+S+S'+l+J_f$ and the overlap integral is
given as
\begin{eqnarray}
I_{if}= \sqrt{\frac {3}{4 \pi}} \int^{\infty}_{0}
u_{E}^{(lSJ)}(r)u^{(l_fS'J_f)} (r) dr.
\end{eqnarray}
In the above equations the magnetic momenta are taken as $\mu_{p}=$2.792847
$\mu_N$ and $\mu_{^7\rm Be}$=-1.398 $\mu_N$ for the first and second
particles, respectively.

Finally, the astrophysical $S$ factor of the process is expressed in
terms of the cross section with the help of the equation
\cite{Fowler}
\begin{eqnarray}
S(E)=E \, \, \sigma(E) \exp(2 \pi \eta).
\end{eqnarray}

\section{Numerical results}
\subsection{Details of the calculations and interaction potentials}

\par The Schr\"{o}dinger equation in the entrance
and exit channels is solved with the two-body $p-^{7}${\rm Be}
central potentials of the Gaussian form \cite{dub19} as defined in
Eq.(\ref{pot}) with the corresponding point-like Coulomb part. For
consistency we use the same model parameters as in the
aforementioned paper: $\hbar^2/2$[a.m.u]=20.7343 MeV fm$^2$, $m_{\rm
p}=A_1$ a.m.u. $=$1.0072764669 a.m.u., $m_{^7{\rm Be}}=A_2$ a.m.u.
$=$ 7.014735 a.m.u.
\begin{table}[htbp]
\caption{Values of the depth ($V_0$) and width ($\alpha_0$)
parameters of the original and modified $p-^7${\rm Be} potentials
$\textrm{V}_\textrm{D}$ and $\textrm{V}_\textrm{M}$ in different
partial waves.} {\begin{tabular}{@{}ccccc@{}} \toprule $^{2S+1}L_J$
& $V_0$, (MeV) & $\alpha_0$, (fm$^{-2}$) & E$^{^8\rm
B}_{\textrm{FS}}$, (MeV)
\\\colrule
$^3S_1$ & -343.0 & 1.0 & -110.13 \\
$^3S_1$($\textrm{V}_\textrm{M}$) & -100.0 & 0.876 & -2.42 \\
$^3P_0$ & -580.0 & 1.0 & -102.25  \\
$^3P_1$ & -709.85 & 0.83 & -205.38  \\
$^3P_2$ & -330.414634 & 0.375 & -96.59  \\
$^3P_2$($\textrm{V}_\textrm{M}$) & -300.5003 & 0.34 & -87.86 \\
$^3D_1$ & -343.0 & 1.0 & -  \\
$^3D_2$ & -116.04 & 0.095 & -20.45  \\
$^3D_2$($\textrm{V}_\textrm{M}$) & -193.0 & 0.15 & -37.92 \\
$^3D_3$ & -343.0 & 1.0 & -  \\
$^3F_3$ & -104.555 & 0.055 & -15.99  \\\botrule
\end{tabular}\label{tab1}}
\end{table}
The scattering wave function $u_{E}(r)$ of the relative motion is
obtained by solving the Schr{\"o}dinger equation using the Numerov
method with an appropriate potential subject to the boundary
condition specified in Eq.(\ref{eq220}).

The depth $V_0$ and width $\alpha_0$ of the $p-^7${\rm Be}
potentials are given in Table \ref{tab1}. We use two parameter sets
for the original potential $V_D$ from Ref. \cite{dub19} and the
modified potential $V_M$, respectively. These potentials differ from
each other only in the $^3S_1$, $^3P_2$ and $^3D_2$ partial waves.
The last column of the table contains energies of the forbidden
states in the $^3S_1$, $^3P_1$, $^3P_2$, $^3P_3$, $^3D_2$ partial
waves. The parameters of the modified $V_M$ potential are fitted to
reproduce the scattering length $a_{01}$ in the $^3S_1$ partial
wave, binding energy of the $^8{\rm B}(2^+,1)$ ground state and the
experimental astrophysical $S$ factor at the Gamow energy in the
$^3P_2$ partial wave, and the experimental astrophysical $S$
factor around the $^3D_2$ resonance.

First we examine how the scattering length $a_{01}$ is described
with the original $V_D$ potential in the $^3S_1$ wave. As discussed
in the Introduction, the most realistic experimental data
$a_{01}^{exp}=17.34^{+1.11}_{-1.33}$ fm \cite{pan19} for the spin=1
channel should be reproduced by the $p-^7\rm Be$ potential. However,
the original $V_D$ potential yields an estimate of
$a_{01}^{th}$=-0.26 fm, which does not reproduce even the sign of
the data. In Table \ref{tab1} we present the fitted parameters of
the new modified potential $V_M$ in the $^3S_1$ partial wave which
yields an estimate of $a_{01}^{th}$=17.34 fm for the scattering
length. 
The parameters of the modified potential $V_M$ in the $^3P_2$ bound
channel are adjusted according to two conditions. The first
condition for the potential is the binding energy $E_b$=0.1375 MeV
of the $^8{\rm B}(2^+,1)$ ground state. The second condition comes
from Eq.(\ref{eq3}). It represents the experimental value of the
astrophysical $S$ factor at the Gamow solar energy. 
The last condition could as well be replaced by the relation
 \beq S_s(0)/C^2\approx 35.6(1-0.0014 a_{01}) \, {\rm eV \, b \,fm} \approx 34.74  \, {\rm eV \, b \,fm},
 \label{eq21}
 \eeq
from Ref. \cite{baye00b}. This relationship connects the scattering
length with the ANC and the astrophysical $S$ factor at zero energy
due to the transition from the initial $S$ scattering wave. In other
words, the above two conditions from Eq.~(\ref{eq3}) and
Eq.~(\ref{eq21}) should be equivalent. Since Eq.~(\ref{eq21})
needs extrapolated value of the astrophysical $S$ factor at $E=0$,
its uncertainty is quite large. This is why we use Eq.~(\ref{eq3})
to define the potential parameters in the $^3P_2$ 
partial wave. Then consistency of the new potential with the
relation in Eq. (\ref{eq21}) will be examined. The original $V_D$
and the modified $V_M$ potentials yield values $C^2=0.496$ fm$^{-1}$
and $C^2=0.538$ fm$^{-1}$, respectively, for the ANC of the bound
$^3P_2$ state. 
\begin{figure}[htbp]
\includegraphics[width=
8.4cm
]{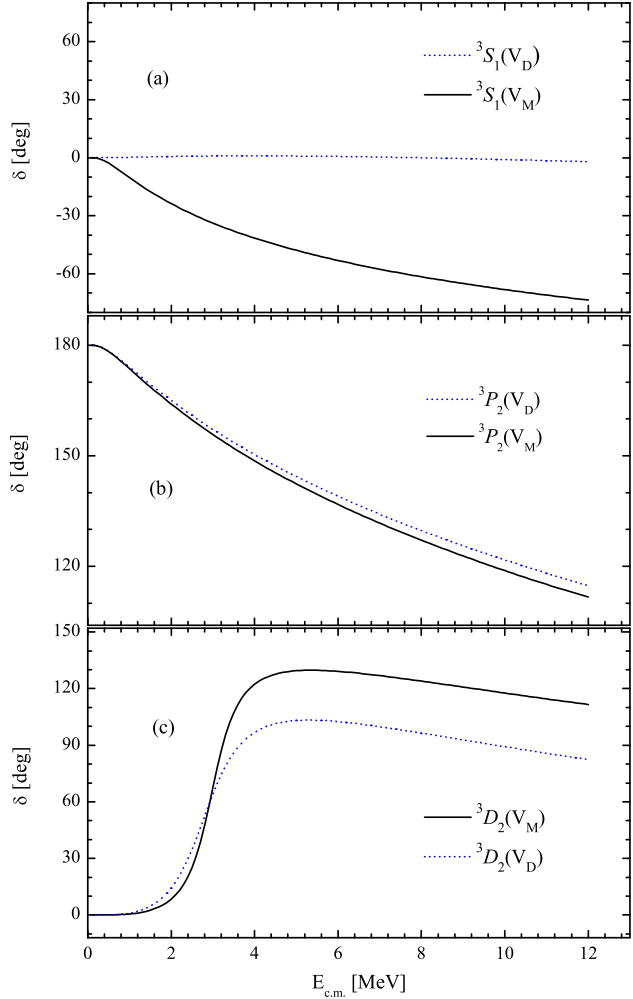}
\caption{Phase shifts in the $^3S_1$ (a), $^3P_2$ (b) and $^3D_2$
(c) partial waves of the $p-^7${\rm Be} scattering state with
potentials $\textrm{V}_\textrm{D}$ and $\textrm{V}_\textrm{M}$.}
\label{f1}
\end{figure}
Finally, the parameters of the modified potential in the partial
$^3D_2$ wave are chosen to reproduce the astrophysical $S$ factor
in the second resonance region around $E=3$ MeV.

In Fig. \ref{f1} we show the description of the phase shift in the
$^3S_1$, $^3P_2$ and $^3D_2$ partial waves. As can be seen from the
figure, the potentials $V_D$ and $V_M$ yield similar phase-shift
description in the partial waves $^3P_2$ and $^3D_2$, but display
significantly different description in the partial $^3S_1$ wave channel.

\subsection{
The astrophysical $S$ factor and the reaction rates
of the $^{7}{\rm Be}(p, \gamma)^{8}{\rm B}$ capture process}

\begin{figure}[htbp]
\includegraphics[width=8.4cm]{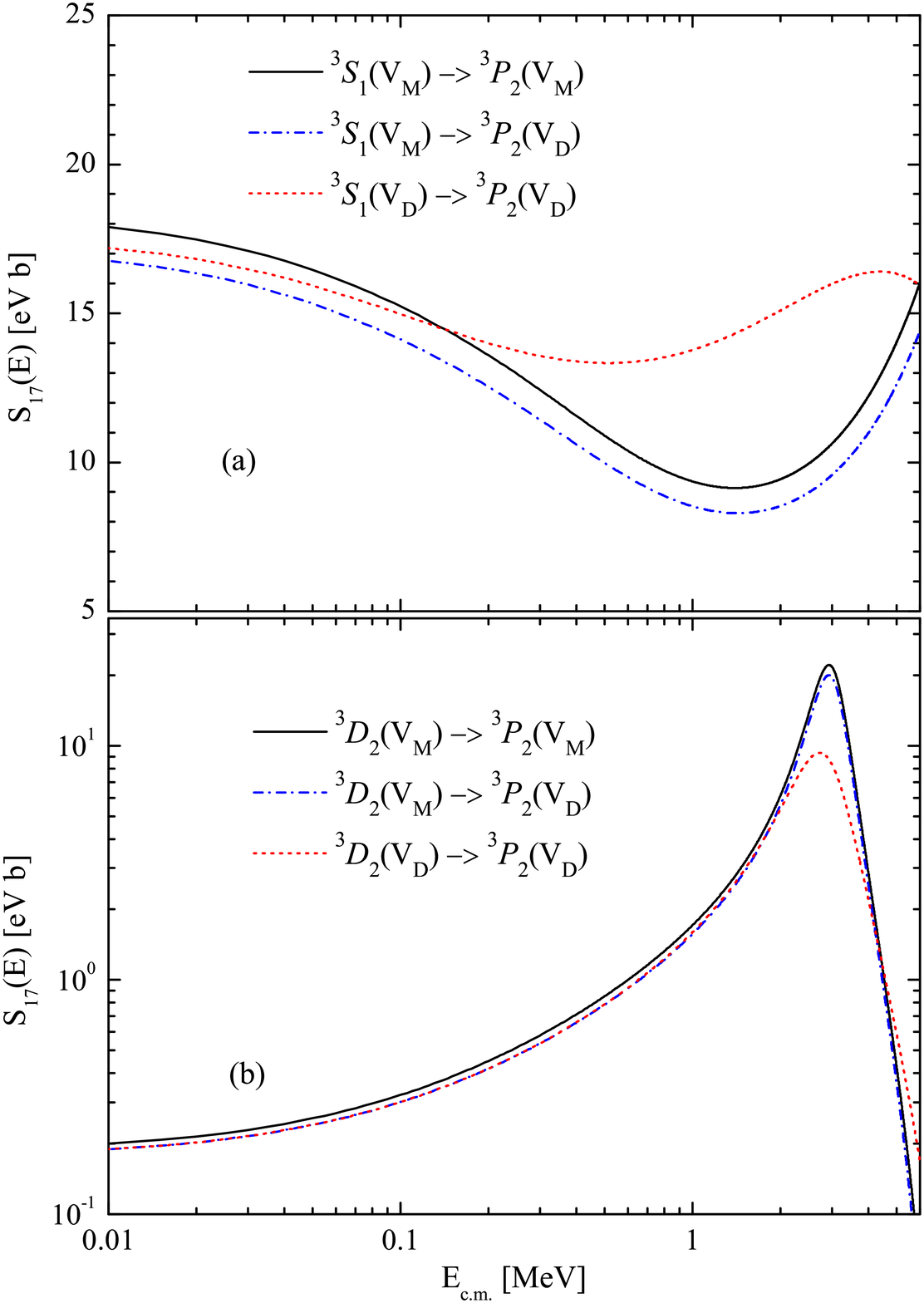}
\caption{Astrophysical $S$ factors for the $^{7}{\rm Be}(p,
\gamma)^{8}{\rm B}$  synthesis reaction due to the E1 transitions $^3S_1
\to ^3P_2$ (a) and $^3D_2 \to ^3P_2$ (b) estimated within the
potential models $V_D$ and $V_M$, and their combination.} \label{f2}
\end{figure}

The astrophysical $S$ factor and reaction rates of the $^{7}{\rm
Be}(p, \gamma)^{8}{\rm B}$ direct radiative-capture process
presented below are calculated with the potentials $V_D$ and
$V_{M}$. The partial astrophysical $S$ factors estimated with the
above potential models and their combination for the initial $^3S_1$
channel are presented in Fig. \ref{f2} (panel a). As can be seen,
the potential model $V_M$ yields results quite different from the
$V_D$ model ones for both absolute values and energy dependence of
the $S$ factor. This is, firstly, due to the fact that these models
yield different values for the scattering length $a_{01}$ and,
secondly, due to the relation between the astrophysical $S$ factor
and the scattering length $a_{01}$ given in Eq. (\ref{eq21}). On the
other hand, the value of $S_s(0.6 \rm keV)/C^2=35.18 \, \rm eV \, b
\,fm$, calculated for the $^3S_1 (V_M) \to ^3P_2 (V_D)$ transition
with a combined potential model is larger than the value of 34.74 eV
b fm from Eq.(\ref{eq21}). The corresponding estimate for the
$^3S_1(V_M) \to ^3P_2(V_M)$ transition at the energy $E=0.6$ keV is
about 34.57 eV b fm, which is more consistent with the underlying
theory \cite{baye00b}.

In panel (b) of Fig. \ref{f2} we show the partial astrophysical $S$
factors estimated for the initial $^3D_2$ resonance channel. Here
the parameters of the model $V_M$ have been adjusted to reproduce
the experimental astrophysical $S$ factor around the resonance
energy. Below we see that this is possible.

\begin{figure}[htbp]
\includegraphics[width=8.4cm]{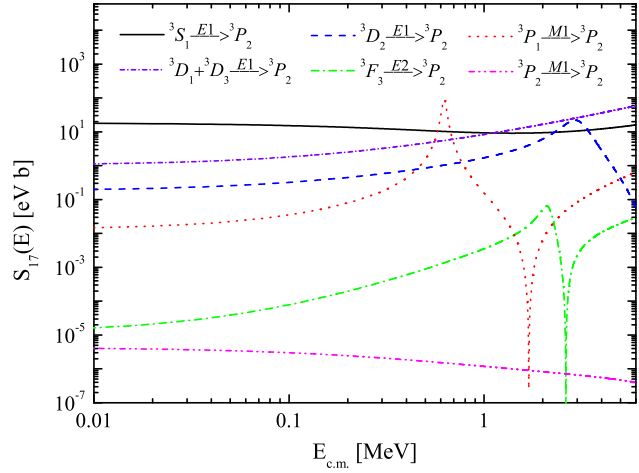}
\caption{The partial E1, E2  and M1 components of the
astrophysical $S$ factor for the $^{7}{\rm Be}(p, \gamma)^{8}{\rm
B}$ capture process within the $V_{M}$ potential model.} \label{f3}
\end{figure}

Figure \ref{f3} compares the partial astrophysical $S$ factors for
different initial scattering channels obtained within the potential
model $V_M$. One can see that the most important contribution at low
energies comes from the initial $^3S_1$ channel due to the electric
E1 transition. The E1 transitions from the initial $^3D_1$,
$^3D_2$ and $^3D_3$ scattering channels altogether yield a
contribution that is less than the contribution from the main
$^3S_1$ channel by an order of magnitude at low energies. However,
they become comparable at energies beyond the resonance region. The
partial M1 transition from the initial $^3P_1$ scattering wave and
E1 transition from the $^3D_2$ wave are responsible for the first
and second resonances at energies 0.633 MeV and 2.988 MeV,
respectively.

\begin{figure}[htbp]
\includegraphics[width=8.4cm]{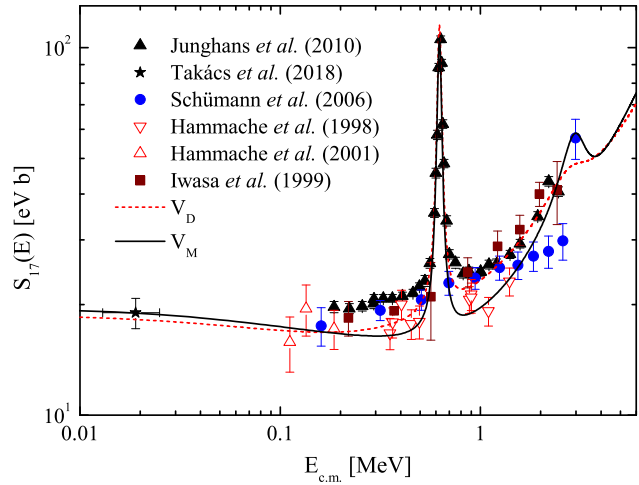}
\caption{Astrophysical $S$ factor for the $^{7}{\rm Be}(p,
\gamma)^{8}{\rm B}$  synthesis reaction within the potential models
$V_D$ and $V_M$ in comparison with available experimental data.}
\label{f4}
\end{figure}

In Fig.\ref{f4} we present the total astrophysical $S$ factor of the
$^{7}{\rm Be}(p, \gamma)^{8}{\rm B}$ process obtained within the
potential models $V_D$ and $V_M$. As can be seen from the figure,
the results for the potential model $V_M$ are mostly consistent with
the two data sets of Hammache {\it et al.} \cite{ham98,ham01}. Other
measurements \cite{jun10,tak18,schum06,iwa99} show higher values in
the vicinity of the resonance.

A behavior of the astrophysical $S$ factor near zero energy is more
complex. Our estimates within the $V_M$ potential model are
$S_{17}(\rm 1 \, keV) =19.64$ eV b and $S_{17}(\rm 0.6\, keV)=20.07$
eV b. A linear extrapolation to the zero energy yields \beq
S_{17}(0) \approx (20.5 \pm 0.5) \, \rm eV \, b ,
 \eeq
which is consistent with the SF II estimate \cite{adel11} quoted in
Eq.(\ref{eq01}).

\begin{figure}[htbp]
\includegraphics[width=8.4cm]{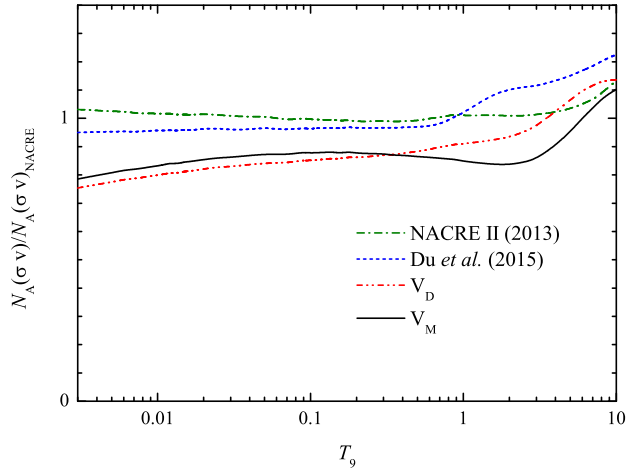}
\caption{Reaction rates of the direct $p+^7\rm Be \rightarrow
^8$B$+\gamma$ capture process within the $V_D$ and $V_M$ potential
models normalized to the experimental data by the NACRE
collaboration \cite{ang99}} \label{f5}
\end{figure}

Finally, estimated reaction rates within the models $V_D$ and $V_M$
are presented in Table \ref{tab02} and Fig. \ref{f5}. In the second
and third columns of the table, "the most effective" energy $E_0$
and the width of the Gamow window $\Delta E_0$ are given
\cite{ang99}. One can note that our theoretical results are
substantially lower than the estimates of the NACRE II collaboration
\cite{xu13} and Du {\it et al.} \cite{du15}.

\begin{table*}[htbp]
\caption{Theoretical estimations of the direct $^{7}{\rm Be}(p,
\gamma)^{8}{\rm B}$ capture reaction rate in the temperature
interval $10^{6}$ K $\leq T \leq 10^{10}$ K ($ 0.001\leq T_{9} \leq
10 $).} \label{tab02}

\begin{center}
{\scriptsize
\begin{tabular}{ccccccccccc} \hline \hline $T_9$ &
$E_0$ (MeV) & $\Delta E_0$ (MeV) & \multicolumn{2}{c}{$N_{A}(\sigma
v)$ ($ \textrm{cm}^{3} \rm{mol}^{-1} \rm{s}^{-1}$)}
&$\,\,\,\,\,\,\,$& $T_9$ & $E_0$ (MeV) & $ \Delta E_0$ (MeV) &
\multicolumn{2}{c}{$N_{A}(\sigma v)$ ($ \textrm{cm}^{3}
\rm{mol}^{-1} \rm{s}^{-1}$)} \\ 
 &  &  & $\textrm{V}_\textrm{D}$ & $\textrm{V}_\textrm{M}$ &  &&  &  & $\textrm{V}_\textrm{D}$ & $\textrm{V}_\textrm{M}$\\
 \hline
0.001 & 0.003 & 0.001 & $4.99\times10^{-38}$ & $5.19\times10^{-38}$ && 0.120 & 0.072 & 0.063 & $8.15\times10^{-4}$ & $8.39\times10^{-4}$\\
0.002 & 0.005 & 0.002 & $5.19\times10^{-29}$ & $5.41\times10^{-29}$ && 0.130 & 0.076 & 0.067 & $1.34\times10^{-3}$ & $1.37\times10^{-3}$ \\
0.003 & 0.006 & 0.003 & $1.21\times10^{-24}$ & $1.26\times10^{-24}$ && 0.140 & 0.079 & 0.072 & $2.09\times10^{-3}$ & $2.14\times10^{-3}$\\
0.004 & 0.007 & 0.004 & $6.77\times10^{-22}$ & $7.05\times10^{-22}$ && 0.150 & 0.083 & 0.076 & $3.12\times10^{-3}$ & $3.20\times10^{-3}$  \\
0.005 & 0.009 & 0.004 & $6.07\times10^{-20}$ & $6.32\times10^{-20}$ && 0.160 & 0.087 & 0.080 & $4.51\times10^{-3}$ & $4.61\times10^{-3}$ \\
0.006 & 0.010 & 0.005 & $1.87\times10^{-18}$ & $1.94\times10^{-18}$ && 0.180 & 0.094 & 0.088 & $8.63\times10^{-3}$ & $8.81\times10^{-3}$ \\
0.007 & 0.011 & 0.006 & $2.87\times10^{-17}$ & $2.99\times10^{-17}$ && 0.200 & 0.101 & 0.096 & $1.51\times10^{-2}$ & $1.53\times10^{-2}$ \\
0.008 & 0.012 & 0.007 & $2.72\times10^{-16}$ & $2.84\times10^{-16}$ && 0.250 & 0.117 & 0.116 & $4.58\times10^{-2}$ & $4.63\times10^{-2}$ \\
0.009 & 0.013 & 0.007 & $1.82\times10^{-15}$ & $1.90\times10^{-15}$ && 0.300 & 0.132 & 0.135 & $1.06\times10^{-1}$ & $1.07\times10^{-1}$ \\
0.010 & 0.014 & 0.008 & $9.35\times10^{-15}$ & $9.74\times10^{-15}$ && 0.350 & 0.146 & 0.153 & $2.07\times10^{-1}$ & $2.07\times10^{-1}$ \\
0.011 & 0.015 & 0.009 & $3.90\times10^{-14}$ & $4.06\times10^{-14}$ && 0.400 & 0.160 & 0.172 & $3.60\times10^{-1}$ & $3.57\times10^{-1}$ \\
0.012 & 0.015 & 0.009 & $1.38\times10^{-13}$ & $1.44\times10^{-13}$ && 0.500 & 0.186 & 0.207 & $8.53\times10^{-1}$ & $8.35\times10^{-1}$ \\
0.013 & 0.016 & 0.010 & $4.27\times10^{-13}$ & $4.45\times10^{-13}$ && 0.600 & 0.210 & 0.240 & $1.66\times10^{0}$ & $1.60\times10^{0}$ \\
0.014 & 0.017 & 0.010 & $1.18\times10^{-12}$ & $1.23\times10^{-12}$ && 0.700 & 0.232 & 0.273 & $2.85\times10^{0}$ & $2.72\times10^{0}$ \\
0.015 & 0.018 & 0.011 & $2.97\times10^{-11}$ & $3.10\times10^{-12}$ && 0.800 & 0.254 & 0.306 & $4.50\times10^{0}$ & $4.27\times10^{0}$ \\
0.016 & 0.019 & 0.012 & $6.92\times10^{-12}$ & $7.20\times10^{-12}$ && 0.900 & 0.275 & 0.337 & $6.69\times10^{0}$ & $6.28\times10^{0}$ \\
0.018 & 0.020 & 0.013 & $3.07\times10^{-11}$ & $3.20\times10^{-11}$ && 1.000 & 0.295 & 0.368 & $9.46\times10^{0}$ & $8.82\times10^{0}$ \\
0.020 & 0.022 & 0.014 & $1.11\times10^{-10}$ & $1.15\times10^{-10}$  && 1.500 & 0.386 & 0.516 & $3.18\times10^{1}$ & $2.89\times10^{1}$\\
0.025 & 0.025 & 0.017 & $1.44\times10^{-9}$ & $1.50\times10^{-9}$ && 2.000   & 0.468 & 0.656 & $6.45\times10^{1}$ & $5.78\times10^{1}$\\
0.030 & 0.028 & 0.020 & $1.01\times10^{-8}$ & $1.05\times10^{-8}$    && 2.500 & 0.543 & 0.790 & $1.03\times10^{2}$ & $9.13\times10^{1}$ \\
0.040 & 0.034 & 0.025 & $1.72\times10^{-7}$ & $1.78\times10^{-7}$    && 3.000 & 0.613 & 0.919 & $1.45\times10^{2}$ & $1.28\times10^{2}$\\
0.050 & 0.040 & 0.030 & $1.27\times10^{-6}$ & $1.32\times10^{-6}$    && 4.000 & 0.743 & 1.168 & $2.38\times10^{2}$ & $2.10\times10^{2}$\\
0.060 & 0.045 & 0.035 & $5.82\times10^{-6}$ & $6.02\times10^{-6}$    && 5.000 & 0.862 & 1.407 & $3.41\times10^{2}$ & $3.05\times10^{2}$\\
0.070 & 0.050 & 0.040 & $1.95\times10^{-5}$ & $2.02\times10^{-5}$    && 6.000 & 0.973 & 1.638 & $4.52\times10^{2}$ & $4.12\times10^{2}$\\
0.080 & 0.055 & 0.045 & $5.27\times10^{-5}$ & $5.45\times10^{-5}$    && 7.000 & 1.078 & 1.863 & $5.70\times10^{2}$ & $5.30\times10^{2}$\\
0.090 & 0.059 & 0.049 & $1.22\times10^{-4}$ & $1.26\times10^{-4}$    && 8.000 & 1.179 & 2.082 & $6.93\times10^{2}$ & $6.55\times10^{2}$\\
0.100 & 0.063 & 0.054 & $2.50\times10^{-4}$ & $2.58\times10^{-4}$    && 9.000 & 1.275 & 2.297 & $8.18\times10^{2}$ & $7.84\times10^{2}$\\
0.110 & 0.068 & 0.058 & $4.69\times10^{-4}$ & $4.83\times10^{-4}$    && 10.00 & 1.368 & 2.507 & $9.45\times10^{2}$ & $9.18\times10^{2}$ \\
\hline \hline
\end{tabular}
}
\end{center}
\end{table*}

\section{Conclusions}

The astrophysical $^{7}{\rm Be}(p, \gamma)^{8}{\rm B}$ direct
capture process has been studied within the two-body potential model
using the single-channel approximation.
The modified potential is
constructed to reproduce the new  experimental value of the $S$-wave
scattering length and the known astrophysical $S$ factor at the
Gamow energy, extracted from the solar neutrino flux. The modified
potential is consistent with the theory of Baye \cite{baye00a}
which connects the $S$-wave scattering length with the astrophysical
$S$ factor at zero energy divided by the square of ANC.

The results
obtained for the astrophysical $S$ factor within the
modified potential approach are in accordance with the data of Hammache
{\it et al.} in contrast to those obtained using the original potential by Dubovichenko {\it et al.}
 \cite{dub19}. The value of the astrophysical $S$ factor
extrapolated to zero energy is found to be $ S_{17}(0) \approx
(20.5 \pm 0.5) \, \rm eV \, b $ which is consistent with the SF II
estimates \cite{adel11}. However, the calculated reaction rates are
lower than the results of the NACRE II collaboration \cite{xu13}.

\section*{Acknowledgements}
The authors thank Daniel Baye for stimulating discussions of the
problem.  A.S.K. acknowledges the support from the Australian
Research Council. L.D.B. acknowledges the support by the Russian
Science Foundation for Basic Research Grant No. 19-02-00014.

\end{document}